\documentclass[aps,prd,letterpaper,reprint,superscriptaddress,showpacs,preprintnumbers,nofootinbib,10pt]{revtex4-1}
\usepackage{graphicx}  
\usepackage{subcaption}
\usepackage{dcolumn}   
\usepackage{bm}         
\usepackage{amssymb}   
\usepackage{amsmath}
\usepackage{mathrsfs}  
\usepackage{float}
\usepackage{graphicx,color} 
\usepackage{ragged2e}
\usepackage{caption}
\usepackage[english]{babel}     
\usepackage{graphicx,wrapfig,lipsum}      
\usepackage{multimedia}    
\usepackage[mathscr]{eucal}  
\usepackage{bm}              
\usepackage{amssymb}        
\usepackage{amsmath}     
\usepackage{mathrsfs}  
\usepackage[mathscr]{eucal}   
\usepackage{subcaption}

\bmdefine\va{a}
\bmdefine\vb{b}
\bmdefine\vc{c}
\bmdefine\ve{e}
\bmdefine\vf{f}
\bmdefine\vj{j}
\bmdefine\vk{k}
\bmdefine\vn{n}
\bmdefine\vp{p}
\bmdefine\vq{q}
\bmdefine\vr{r}
\bmdefine\vs{s}
\bmdefine\vt{t}
\bmdefine\vv{v}
\bmdefine\vw{w}
\bmdefine\vx{x}
\bmdefine\vy{y}
\bmdefine\vz{z}
\bmdefine\vA{A}
\bmdefine\vB{B}
\bmdefine\vC{C}
\bmdefine\vD{D}
\bmdefine\vE{E}
\bmdefine\vF{F}
\bmdefine\vH{H}
\bmdefine\vI{I}
\bmdefine\vJ{J}
\bmdefine\vK{K}
\bmdefine\vL{L}
\bmdefine\vM{M}
\bmdefine\vO{O}
\bmdefine\vP{P}
\bmdefine\vR{R}
\bmdefine\hvR{\hat{R}}
\bmdefine\vS{S}
\bmdefine\vT{T}
\bmdefine\vW{W}
\bmdefine\vX{X}
\bmdefine\vV{V}
\bmdefine\vN{N}
\bmdefine\vu{u}
   
 \allowdisplaybreaks
\begin{document}     

\title{The ’t Hooft loop from a center-vortex wave functional}

\author{D. R. Junior}
\affiliation{
Instituto de F\'isica Te\'orica, Universidade Estadual Paulista and South-American Institute of Fundamental Research ICTP-SAIFR, Rua Dr. Bento Teobaldo Ferraz, 271 - Bloco II, 01140-070 S\~ao Paulo, SP, Brazil}
\affiliation{Institut f\"ur Theoretische Physik, Universit\"at T\"ubingen, Auf der Morgenstelle 14, 72076 T\"ubingen, Germany}
\author{L. E. Oxman}
\affiliation{
Instituto de F\'isica, Universidade Federal Fluminense, 24210-346 Niter\'oi, RJ, Brasil.}  
\author{H. Reinhardt}

\affiliation{Institut f\"ur Theoretische Physik, Universit\"at T\"ubingen, Auf der Morgenstelle 14, 72076 T\"ubingen, Germany}

\date{\today}

\begin{abstract}     
 
Previously, we proposed an infrared vacuum wave functional for $SU(N)$ Yang-Mills theory peaked at thin center vortices and showed that it yields an area law for the Wilson loop. In this work, we use this wave functional to calculate the spatial 't Hooft loop, for which we find a perimeter law, in accordance with 't Hooft's criterion for confinement.

\end{abstract}

\maketitle     

\section{Introduction}

 To probe confinement, the most common order parameter is the Wilson loop $W(C)$ \cite{wilson1974}, which is given by the holonomy of the gauge field along a closed curve $C$. An alternative order parameter, which can be thought of as the dual to the Wilson loop, was introduced by 't Hooft \cite{thooft} and is commonly referred to as 't Hooft loop. These two order parameters provide complementary observables for studying the Yang-Mills vacuum. In a confining phase, the Wilson loop obeys an area law, while the 't Hooft loop satisfies a perimeter law. In contrast, in a Higgs phase, these behaviors are reversed. 

Any theoretical framework aimed at understanding confinement must provide a unified explanation for these distinct behaviors. 

 The 't Hooft loop operator $V(C_1)$ of a general closed curve $C_1$ was introduced by the algebraic equation \cite{thooft}
\begin{align}
	V(C_1)W(C_2)=Z^{L(C_1,C_2)}W(C_2)V(C_1)\;,
\end{align}
where  $W(C_2)$ is the Wilson loop in the defining representation, $Z=e^{-i\frac{2\pi}{N}}$ is a center element, and $L(C_1,C_2)$ is the Gauss' linking number. The 't Hooft loop was evaluated on the lattice for $SU(2)$ in Refs. \cite{thooftlattice1,thooftlattice2}. For the spatial 't Hooft loop a perimeter law was found at zero temperature, while for high temperatures an area law was observed.

An explicit representation of the 't Hooft loop operator in continuum Yang-Mills theory was obtained in Ref.  \cite{thooftloop} and is given by
\begin{align}
	V(C)=\exp\left[i\int d^3x\, a(x,C) \cdot\Pi(x)\right]\;,\label{thooftexplicit}
\end{align}
where $a(C)$ is the gauge field of a thin center vortex located on the loop $C$, whose explicit form is given in Eq. ~\eqref{a2}. Furthermore $\Pi(x)$ is the canonical momentum operator, which in the coordinate (gauge field) representation reads:
\begin{align}
	\label{thooft-op}
	\Pi_k^a(x)=-i\frac{\delta}{\delta A_k^a(x)} \;.
\end{align}
In this representation, the operator \eqref{thooftexplicit} translates the argument of a wave functional by the center-vortex field $a(C)$, i.e.
\begin{align}
	\label{vort-gen}
V(C)\Psi(A)=\Psi(A+a(C))\;.
\end{align}
In this sense the 't Hooft loop operator $V(C)$ creates a center-vortex loop with guiding center $C$.
While the Wilson loop $W(C)$ depends only on the field coordinate (i.e. the gauge field) $A(x)$, the 't Hooft loop operator $V(C)$ depends only on the canonical conjugate momentum, thereby making their duality manifest.
With the representation \eqref{thooftexplicit}, the ’t~Hooft loop was computed in the confinement regime in Ref.~\cite{reinhardt2007} using the vacuum
wave functional obtained in the variational approach in Coulomb gauge~\cite{reinhardt2005}, and a perimeter law was found.

Among the proposed confinement mechanisms,
the center vortex picture stands out as a promising candidate.
This picture, introduced already in the late 70s \cite{basic-v1,basic-v2,HBN}, received a revival at the beginning of this century through lattice calculations carried out in the so-called maximal center gauge \cite{det2}, where one fixes only the coset $G/Z$ but leaves the center $Z$ of the gauge group $G$ unfixed. After center projection, which replaces each link by its closest center element and thereby extracts the vortex content of the gauge field, virtually the full string tension is obtained \cite{det2,PhysRevD.69.014503}. Furthermore, the density of the so extracted center vortices shows the proper scaling towards the continuum limit, indicating that these vortices are physical objects \cite{basic-v4}.
On the other hand, removing center vortices in the lattice simulations yields a vanishing string tension, restores chiral symmetry \cite{cp2}, and, in accordance with the Banks–Casher theorem, the quark spectrum develops a gap near zero virtuality \cite{Gattnar:2004gx}, see also \cite{chiralsb}. Given the relevance of center vortices for the spontaneous breaking of chiral symmetry, and the fact that this phenomenon dominates low-energy hadron physics, it is not surprising that center vortices also manifest themselves in the hadron spectrum, as observed in Ref.~\cite{cvfermions3}. Removing center vortices also removes the topological charge (Pontryagin index) \cite{cp2}. The topological charge of a center vortex configuration is given by the self-intersection number of the closed vortex surface \cite{top-1,Reinhardt:2001kf}, and alternatively by the temporal change of the writhing number of their three-dimensional projections \cite{Reinhardt:2001kf}.
Furthermore, a nonzero topological charge requires nonoriented center-vortex configurations \cite{Reinhardt:2001kf}, where the flux in the Lie algebra changes orientation at monopole loops. 

The dominance of the center vortices in the Yang-Mills vacuum can be understood from the lattice result that the excitation probability for a sufficiently thick vortex in the vacuum is basically one \cite{thick-latt}. 
This goes along with the observation that the one-loop energy density of a magnetic center vortex configuration is degenerate with that of the perturbative vacuum $A_\mu=0$ \cite{LEH}. 

In the center vortex picture the deconfinement phase transition shows up as a depercolation transition from a confined phase of percolated vortices to a gas of smoothly interacting small vortices winding dominantly around the compactified Euclidean time axis \cite{vor-b6}. Furthermore it was found first in a random vortex model \cite{cv-branching} and subsequently in lattice calculations \cite{cv-branching-1}, \cite{vor-branch} that the deconfinement phase transition is accompanied by a substantial reduction of the vortex branching, which occurs in $SU(N>2)$ gauge theory due to the existence of several nontrivial center elements. 

In Ref. \cite{ox4d}, it was proposed that a mixed ensemble of oriented and nonoriented center vortices with non-Abelian degrees of freedom in 4D can describe $N$-ality together with the formation of a soliton-like confining flux tube. The obtained effective description was analyzed in Refs. \cite{o-verc,oxgustavo,stability,prospecting}, showing that it can accommodate Abelian-like transverse energy profiles, a Casimir law for asymptotic string tensions, as well as possible deviations. Indeed, in lattice measurements, the asymptotic profiles \cite{fluxtube-1,fluxtube-2,fluxtube-3} and string tensions \cite{teper,latt-scaling} are close to these behaviors.\footnote{Note that the asymptotic Casimir law refers to distances where $N$-ality is settled, and not to the intermediate confining region where Casimir scaling was established \cite{casimir-1}, which can be explained as due to the effect of center-vortex thickness \cite{thick-1,thick-2}.} The mixed ensemble was also analyzed in 3D Euclidean spacetime \cite{dgo}. In this case, since vortices generate worldlines, they are described by effective complex fields \cite{stone,samuel-1,samuel-2}. When the parameters correspond to a condensate, the model leads to the formation of a soliton-like domain wall sitting on the Wilson loop, whose asymptotic behavior reproduces an asymptotic Casimir law, in agreement with lattice simulations \cite{LT}. In these works, the partition functions in both 4D and 3D Euclidean spacetime were constructed to capture the simplest correlations among elementary center-vortex worldsurfaces and worldlines, respectively (for a review, see Ref. \cite{universe}).

Stimulated by the success of the center vortex picture, in Ref.  \cite{wavefunctional} we proposed a vacuum wave functional for the infrared regime of $4$D Yang-Mills theory, which is peaked at Abelian-projected center-vortex configurations. The use of the Hamiltonian approach, as compared to the four-dimensional functional integral, has the advantage that only gauge field configurations $A(x)$ at a given time-slice $x \in \mathbb{R}^3$ enter the wave functional $\Psi[A]$.  In this manner, the configurations become networks of oriented and nonoriented center-vortex lines with $N$-matching, which are easier to handle than the two-dimensional vortex worldsurfaces in $D=4$. With this wave functional, we calculated the Wilson loop and obtained an area law that also exhibits Casimir scaling, through a mechanism that is technically similar to that implemented in 3D spacetime, but pertains to the infrared sector of the $3+1$ dimensional theory. \footnote{Recently, the Abelian-projected ensemble was formulated at the level of the 4D Euclidean partition function by using the Weingarten matrix representation for the sum over surfaces \cite{weingarten}. In that reference, the effective 4D description of the center-vortex ensemble with non-Abelian degrees of freedom \cite{ox4d} was also reexamined and compared with 
 the Abelian-projected one.}

In this work, we will continue this line of research by calculating the 
 't Hooft loop with the center vortex wave functional.

 The structure of the paper is as follows: In Section II, we review the infrared wave functional peaked at center vortices, as constructed in Ref.  \cite{wavefunctional}. Section III provides a brief review of the Wilson loop calculation using our vortex wave functional, in order to demonstrate that the 't Hooft loop computation presented in Section IV is carried out consistently with that of the Wilson loop. Finally, Section V contains our conclusions.

\section{Vacuum wave functional peaked at center vortices}\label{vortexsection}

In $D=3$, an oriented center vortex is a gauge field configuration $a(x,\gamma)$ associated to a closed loop  $\gamma$ whose Wilson loop $W[a(\gamma)](C)$ yields  a nontrivial center element $Z$ of the gauge group provided the two loops are nontrivially linked:
\begin{equation}
	\label{w-loop}
	W[a(\gamma)](C)=Z_m^{L(\gamma,C)} \qquad Z_m =  e^{- i\frac{2\pi m}{N}} \, .
\end{equation}
Here  $L(\gamma,C)$ is the Gauss' linking number and $m=1,2, \dots , N-1 $.
Since $Z_m=Z^m$, $Z =  e^{- i\frac{2\pi}{N}}$ center vortices can split or fuse, see Ref.  \cite{cv-branching}. 

 The gauge-field associated with a thin center-vortex line $\gamma$ is
\begin{align}
	\label{a2}
	a(x,\gamma)&=(-\Delta)^{-1} \nabla \times j(x,\gamma), 
	\\
   j(x,\gamma)&=\mathscr{C}_\gamma \int_{\gamma} d\bar{x}\; \delta(x-\bar{x})\;.
    \label{j2}
\end{align}
Here, $\mathscr{C}_\gamma $ denotes the ``co-weight'' $\mathscr{C}$ of the vortex line $\gamma$ \footnote{Note,  in deviation from the notation of Ref.  \cite{wavefunctional} we have included here the center charge $\mathscr{C}_\gamma $ in the definition of the line current \eqref{j2}.}, which lives in the Cartan subalgebra of $\mathfrak{su}(N)$ with generators $T_q$, $q=1,\dots,N-1$, normalized to
\begin{align}
	{\rm Tr}(T_q T_p) = \frac{1}{2}\delta_{qp}\;.
\end{align} 
In Ref. \cite{wavefunctional}, the ensemble was parametrized in terms of elementary center vortices, which carry center charge $m=1$. The associated fluxes satisfy
\begin{equation}
 e^{i \mathscr{C}}=  Z  \, I \makebox[.5in]{,} Z =  e^{- i\frac{2\pi}{N}}  \;,
 \label{z1}
\end{equation}
where $I$ is the $N \times N$ identity matrix.
There are in fact $N$ different elementary co-weights $\mathscr{C}_k$ satisfying Eq. \eqref{z1} and that furthermore add up to zero
 \begin{equation} 
 	\sum_{k=1}^N \mathscr{C}_{k}=0 \;.
 	\label{w2}
 \end{equation}
 Due to the last property, there are configurations formed by $N$ elementary center-vortex lines $\gamma_k$, which all start at the same point in space and also terminate at the same point, however, each carries a different co-weight $\mathscr{C}_k$. We refer to such vortex configurations as ``$N$-point matchings''. In Ref.  \cite{wavefunctional} we considered center vortex networks $\{\gamma\}$ formed not just by vortex loops but also by $N$-point matchings. Furthermore the networks considered had all possible distributions of elementary co-weights. A wave functional peaked at such vortex networks has the form:
 \begin{align}
 	\label{wave-funct}
    \Psi(A)= \sum_{\{\gamma\}}\psi_{\{\gamma\}}\delta(A- a(\{\gamma\})) \;,
\end{align}
where $a(\{\gamma\})$ is a superposition of the gauge fields $a(x,\gamma)$ \eqref{a2} of all vortex lines $\gamma$ belonging to the vortex cluster $\{\gamma\}$:
\begin{equation}
	a(x,\{\gamma\}) =\sum_{\gamma \in \{\gamma\}} a(x,\gamma) \, .
\end{equation}
The weight factor $\psi_{(\{\gamma\})}$ parametrizes the various properties of the center vortices observed in lattice calculations, such as tension and stiffness, and the sum in Eq.  \eqref{wave-funct} runs over all vortex clusters. 

If $\gamma$ in Eq.  \eqref{a2}  is a closed vortex loop, its magnetic flux is given by
\begin{align}
	\nabla\times a(x,\gamma)= j(x,\gamma)\;. 
	\label{a4}
\end{align}

\begin{figure}
    \centering
    \begin{subfigure}{0.45\linewidth}
    \centering\includegraphics[scale=0.5]{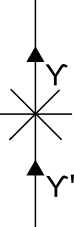}
    \end{subfigure}
     \begin{subfigure}{0.45\linewidth}
    \centering\includegraphics[scale=0.5]{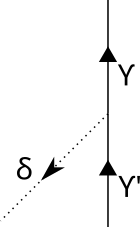}
    \end{subfigure}
     \caption{\justifying Left: chromomagnetic field of a non-oriented configuration formed by two vortex lines $\gamma, \gamma'$ carrying co-weights $\mathscr{C}, \mathscr{C}'$, respectively. In this case, in addition to the flux concentrated along the vortex lines, there is also a component that spreads isotropically around the origin. Right: a non-oriented configuration with a collimated flux. The flux is conserved due to the presence of the unobservable Dirac string $\delta$ with charge $\mathscr{E}=\mathscr{C}'-\mathscr{C}$. }
      \label{fig:collimated}
\end{figure}

In Ref. \cite{wavefunctional} we also included nonoriented center vortices where a pair of vortex lines $\gamma, \gamma'$ carrying different co-weights $\mathscr{C}_{[j]}$, $\mathscr{C}_{[k]}$ meet at a magnetic monopole. (In fact it is the monopole which makes the vortex nonoriented, see Ref.  \cite{Reinhardt:2001kf}.) In this respect, the magnetic flux associated with the gauge field $a(x,\gamma)+a(x,\gamma')$ is partly concentrated on the vortex lines and partly spreads isotropically in space, away from the vortex-line endpoints. However, in our framework, we are interested in describing fluxes which are collimated along the vortex lines, as observed on the lattice (for $SU(2)$, see Ref. \cite{collimated}). For this purpose, unobservable Dirac strings would need to be introduced (see Fig. \ref{fig:collimated}).
Equivalently, these strings can be avoided by describing the nonoriented vortices as non-Abelian objects, which are locally Abelian (see Refs. \cite{conf-qg,ox4d} and \cite{wavefunctional} for further details). In order to avoid unobservable Dirac strings in our Abelian setting, in Ref. \cite{wavefunctional} we introduced a scalar potential $\zeta$, which contributes to the 
magnetic field in the form
\begin{equation}
 B=\nabla \times A -\nabla \zeta \, .
   \label{b1}
\end{equation}
We constructed then a wave functional that is peaked at the possible realizations  of oriented and nonoriented collimated magnetic fluxes and is explicitly given by
\begin{align}
	\Psi
	(A, \zeta)= \sum_{\{\gamma\}}\psi_{\{\gamma\}}\delta(A- a(\{\gamma\}))\delta(\zeta- b(\{\gamma\}))  
	\label{Azeta}. 
\end{align}
Here 
\begin{equation}
	\label{b-mono}
		b(x,\{\gamma\})=(-\Delta)^{-1}\nabla  \cdot \mathcal{B}(x,\{\gamma\})
\end{equation}
and
\begin{gather}
	\mathcal{B}(x,\{\gamma\}) = \sum_{\gamma_n \in \{\gamma\} }j(x,\gamma_n) 
	\label{be}
\end{gather}
is the magnetic flux (c.f. Eq.  \eqref{a4}) of a vortex network $\{\gamma\}$ of vortex lines $\gamma_n$ carrying the Cartan charge $\mathscr{C}_n$. Furthermore the sum in Eq.  \eqref{Azeta} runs over all vortex networks $\{\gamma\}$ including oriented and nonoriented vortices as well as $N$-vortex matching configurations. Note that the scalar monopole potential $\zeta$ enters the wave functional as a dynamical field variable like the gauge field $A$. Consequently, it has to be integrated over in the scalar product of wave functionals.

Due to the presence of the $\delta$-functionals of the gauge field, in our wavefunctional $\Psi(A,\zeta)$ \eqref{Azeta}, it is convenient to switch to a representation in dual variables obtained via functional Fourier transforms:
\begin{align}
    \tilde{\Psi}(  E,  \eta) & =\int [{\mathcal D}A]\int [{\mathcal D}\zeta]\, e^{i \int d^3x\, E\cdot A}  e^{i \int d^3x\, (\eta, \zeta)}  \Psi(A, \zeta) \;,
    \label{we1}
\end{align} 
where  $(A,B)$ is the Killing product in the Lie algebra and $E\cdot A \equiv (E_i,A_i)$.
The dual of the gauge field coordinate $A(x)$ is the electric field $E(x)$, which represents the canonical conjugate momentum, while $\eta$ denotes the dual of the magnetic monopole field $\zeta$. 

Using the definitions of $a(\{\gamma\})$ \eqref{a2} and $b(\{\gamma\})$
\eqref{b-mono}, the dual wave functional in Eq.  \eqref{we1} can be written in the more compact form
\begin{align}
	\tilde{\Psi}(  E,  \eta)  =
	\sum_{\{\gamma\}}\psi_{\{\gamma\}}
	\exp\left(\frac{i}{4\pi} \int_{\{\gamma\}} dx \cdot
 (\Lambda,\mathscr{C}_\gamma)\right) \;.
	\label{we}
\end{align} 
Here, we have introduced the Cartan algebra-valued vector field $\Lambda=\Lambda^q T_q$, which is composed of the dual variables $E,\eta$ and is given by  
\begin{equation}
	\label{lambda}
\Lambda= \Lambda^T+ \Lambda^L  \;,
\end{equation}
with
\begin{align}
&\Lambda^{T}(x)=4\pi\int d^3 \bar{x} D(x-\bar{x})\nabla_{\bar{x}} \times E(\bar{x}) \;,\nonumber\\
&\Lambda^{L}(x)=4\pi\int d^3\bar{x}\,D(x-\bar{x})\nabla_{\bar{x}}\eta(\bar{x})
\;,
\label{LambdaTL}
\end{align}
being its transversal and longitudinal components, respectively. In Eq.  \eqref{we} we have defined
\begin{equation}
	\label{vort-c}
	 \int_{\{\gamma\}} dx\, \cdots =\sum_{\gamma \in\{\gamma\}}  \int_{\gamma} dx \, \cdots \, .
\end{equation}

Model calculations \cite{E-R-2000} show that the probability amplitude $\psi_{\{\gamma\}}$ for the occurrence of a vortex line $\gamma$ or a vortex cluster $\{\gamma\}$ is determined by the vortex tension and stiffness. Parametrizing the vortex lines $\gamma$ by their length $s$, the statistical weight  $\psi_{\{\gamma\}}$  of a vortex network $\{\gamma\}$  is then given by 
\begin{equation}
	\label{stat-weight}
	\psi_{\{\gamma\}}=\exp{\big[-\int_{\{\gamma\}}ds|u(s)|\big(\mu+\dot{u}^2(s)/\kappa\big)\big]}\, ,
\end{equation}
where $u(s)=\dot{x}(s)$ is the tangent vector. The parameters $\mu$ and $\kappa$ control the vortex density and the stiffness, respectively. 
 Furthermore, lattice calculations \cite{cv-corr} show that center vortices have a repulsive interaction. As shown in Ref. \cite{ox-reinhardt}, this can be implemented in our model by introducing an auxiliary scalar field $\sigma(x)$, which contributes a term $-i\sigma$ to the action \eqref{stat-weight}, and has to be integrated over in Eq. \eqref{we} with a Gaussian weight: 
\begin{align}
    \int [D\sigma] e^{-\frac{1}{2\lambda_0}\int d^3x \sigma^2}\;.
\end{align}
Here, $\lambda_0$ measures the strength of the repulsive interactions between the vortex lines weighted by $\lambda_0$. Incorporating this property, together with $N$-matchings and nonoriented components, leads to the same ensemble previously considered in Ref.~\cite{dgo} to study the confining properties of YM theory in $3$D Euclidean spacetime. In that reference (see also Refs. \cite{ox-reinhardt}, \cite{ALB}), using methods of polymer physics \cite{kleinert,fred}, an effective field representation for this ensemble was derived. Then, building on this result, we were able in Ref. \cite{wavefunctional}  to write the dual wave functional $\tilde{\Psi}(E,\eta)$ in terms of the following effective theory for 
a set of $N$ complex scalar fields $\phi_k(x)$: 
\begin{align}
    \tilde{\Psi}(E, \eta) =\prod_{k=1}^N\int[D\bar{\phi}_k][D\phi_k]\, e^{-W(\phi,\Lambda[E,\eta])}\label{infraredansatz} .
\end{align}
Here the index $k$ specifies the $N$ different co-weights $\mathscr{C}_k$ associated to the possible elementary center vortices. Furthermore the ``action" functional for the scalar fields\footnote{Note that $W$ is not an effective action for the 4D ensemble, but rather for the field-representation of its IR vacuum wave functional.} is given by 
\begin{align}
	\label{action}
	&W(\phi,\Lambda)=\int d^3x\left(-\frac{1}{3\kappa}\sum_{k=1}^N\bar{\phi}_k D^2(\Lambda_{k})\phi_k+V(\phi)\right)\;,\\
	&    V(\Phi)=\frac{\lambda_0}{2}\sum_k\left(\bar{\phi}_k\phi_k+\frac{\mu}{\lambda_0}\right)^2-\xi_0\left(\prod_{k=1}^N\phi_k+{\rm c.c.}\right)+\nonumber\\&-\vartheta_0\sum_{k\neq l}\bar{\phi}_k\phi_l\;.\label{fieldtheory}
\end{align}
where 
\begin{equation}
	D(\Lambda_{k})=\nabla-i \Lambda_{k}
\end{equation}
is the covariant derivative of  the Abelian vector field $\Lambda_{k}$ defined by the projections of the Cartan algebra-valued vector field $\Lambda$ \eqref{lambda}  onto the co-weights $\mathscr{C}_k$
\begin{equation}
	\Lambda_{k}=\frac{(\Lambda,\mathscr{C}_k)}{4\pi}\;.
\end{equation}

The parameters $\lambda_0$, $\mu$, and $\kappa$ control the strength of the repulsive interactions between center vortices, the tension, and the stiffness, respectively. In addition, the parameters $\xi_0$ and $\vartheta_0$ determine the relative importance of $N$-matchings and nonoriented configurations. In the absence of these terms, the symmetry of the model would be $U(1)^N$. 
The inclusion of the $\xi_0$ term reduces this symmetry to $U(1)^{N-1}$. 
Finally, the term $\vartheta_0$ associated with non-oriented vortices further breaks the symmetry down to $Z(N)$. In fact, as discussed in Ref. \cite{dgo}, this corresponds to a generalization of 't Hooft's model \cite{thooft-model}, which was formulated with vortices that carry a fixed co-weight $\mathscr{C}$.  Due to this $Z(N)$ symmetry in the percolating regime ($\mu<0$), there are $N$ degenerate vacua corresponding to the $N$ different center elements of $ Z(N)$. 
The equivalence between the representations \eqref{we}, \eqref{infraredansatz} is briefly established in Appendix \ref{app-0}. For the subsequent studies it will be convenient to use the following compact representation for the dual wavefunctional in Eq.  \eqref{infraredansatz}
\begin{align}
	\label{action-sf}
	&\tilde{\Psi}(E,\eta) = \int[D\Phi][D\Phi^\dagger] \exp\left[-W(\Phi,\Phi^\dagger,\Lambda)\right]
	\nonumber \\
	&W(\Phi,\Lambda)=\int d^3x\left({\rm Tr}((D(\Lambda)\Phi)^\dagger (D(\Lambda)\Phi))+V(\Phi,\Phi^\dagger)\right)\;,
	\nonumber\\
	&D(\Lambda)\Phi= \nabla \Phi-i\Lambda\Phi \nonumber \\
	&V(\Phi) = \frac{\lambda}{2}{\rm Tr}(\Phi^\dagger\Phi- a^2I)^2-\xi(\det\Phi+\det\Phi^\dagger)
	\nonumber \\ 
	& 
	-2\vartheta {\rm Tr}(\Phi^\dagger T_A \Phi T_A) \;,
\end{align}
where the field $\Phi$ is an $N\times N$ diagonal matrix with its elements given by the complex scalar fields $\phi_k$. In Eq.  \eqref{action-sf} we have redefined the parameters in the action as follows:
\begin{align}
	&\lambda= \lambda_0 (3\kappa)^{2}, \quad	\xi=\xi_0 (3\kappa)^{N/2}, \quad \vartheta=\vartheta_0 3\kappa,
	\nonumber \\
	&a^2= -\frac{\mu}{(3\kappa)\lambda_0}-\frac{\vartheta}{\lambda}\frac{N-1}{N}=-\frac{1}{3\kappa \lambda_0}(\mu+\vartheta_0 \frac{N-1}{N})
\end{align}

\section{The Wilson Loop}
\label{WL-sect}

In Ref.  \cite{wavefunctional}, using our wave functional \eqref{action-sf}, we evaluated the expectation of the Wilson loop operator 
\begin{align}
	\label{wils-loop-1}
	W(C)= \frac{1}{N}Tr \exp\Big[i\int_C  dx \cdot A(x) \Big]\;.
\end{align}
This was done in the following way:
since the vortex wave functional \eqref{Azeta} has support only on the Cartan gauge potentials \eqref{a2}, \eqref{j2}, the ordinary Stokes' theorem can be used to express the Wilson loop as
\begin{align}
	\label{wils-loop}
	W(C)&= \frac{1}{N} Tr \exp\Big[i\int_{S(C)} d\Sigma\cdot B \Big]
	\nonumber \\
   &=\frac{1}{N} Tr \exp\Big[i\int d^3x\, \Sigma(x,S(C))\cdot B \Big]
\end{align}
where $S(C)$ is an arbitrary surface bordered by $C$ and we have introduced the characteristic function of the surface $S(C)$:
\begin{align}
	\label{charac-fkt}
	\Sigma(x,S(C))=\int_{S(C)}d\sigma_1 d\sigma_2\,\partial_{\sigma_1}x\times \partial_{\sigma_2} x\, \delta(x-\bar{y}(\sigma))\;.
\end{align}
Here $\bar{y}(\sigma)$ is a parametrization of $S(C)$. The surface $S(C)$ can be arbitrarily chosen except for its boundary $\partial S=C$.
Inserting in \eqref{wils-loop} the explicit form of the magnetic field $B$ \eqref{b1} and performing a partial integration we find for the Wilson loop 
\begin{equation}
	\label{wils-loop1}
	W(C)=\frac{1}{N} Tr \exp\Big[i\int d^3x \big[ (\nabla \times  \Sigma)\cdot A + \nabla \cdot \Sigma \, \eta \big] \Big]
\end{equation}
 
In order to perform the trace, it is convenient to use the spectral representation of the Cartan generators $T_q$ 
\begin{equation}
	T_q=\sum_{k=1}^N  \omega_k|_q\, |k\rangle  \langle k|,
\end{equation}
where $|k \rangle$ are the weight vectors and $\omega_k$ is a tuple formed by the eigenvalues $\omega_k|_q$, known as the weights of the defining representation of $SU(N)$.  Using that  $P_k=|k\rangle  \langle k|$ are orthogonal projectors, the trace in the Wilson loop \eqref{wils-loop1} can be transformed to 
\begin{align}
	\label{w-loop2}
	&W(C)=\frac{1}{N} \sum_k  \exp\Big[\frac{i}{4\pi}\int d^3x \big[ (\nabla \times  \Sigma)\cdot (\mathscr{C}_k,A) +\\& \nabla \cdot \Sigma \, (\mathscr{C}_k,\eta) \big] \Big] \;,
\end{align}
where we explicitly used the relation between the co-weights and the defining weights:
\begin{gather}
    \mathscr{C}_k  = 4\pi\omega_k|_q T_q \;.
\end{gather}
In the dual variables $E,\eta$ the gauge and the monopole fields $A, \zeta$ are given by\footnote{Note that $E$ is the negative momentum of the gauge field.} 
\begin{equation}
	A^{q}=i\frac{\delta}{\delta E^{q}}, \qquad 	\zeta^{q}=i\frac{\delta}{\delta \eta^{q}} \;.
\end{equation} 
Using this representation in Eq.  \eqref{w-loop2} the action of the Wilson loop operator on the wave functional in the dual variables yields:
\begin{equation}
	W(C)  \tilde{\Psi}(E,\eta)= \frac{1}{N} \sum_k \tilde{\Psi}(E+\frac{\mathscr{C}_k}{4\pi} \nabla \times  \Sigma ,\eta-\frac{\mathscr{C}_k}{4\pi}\nabla \cdot \Sigma ) \, .
\end{equation}
With this result one obtains for the Wilson loop 
\begin{align}
    &\langle W(C)\rangle=\sum_k \int [DE][D\eta]\, \tilde{\Psi}^*(E,\eta)\times\nonumber \\&\tilde{\Psi}\left(E+\frac{\mathscr{C}_k}{4\pi}\nabla\times \Sigma(S),\eta-\frac{\mathscr{C}_k}{4\pi}\nabla\cdot \Sigma(S) \right)\;.
    \label{p-i}
\end{align}

We first consider the effective theory of the wave function  $\tilde{\Psi}(E,\eta)$. The dominating contribution to the integrals over the $\phi_k$ comes from the vacuum configuration $\phi_k=v$. We can therefore do these integrals in the (lowest order) saddle-point approximation. Upon freezing the scalar fields $\phi_k$ at their vacuum values $v$, the wave functional \eqref{infraredansatz} simplifies to
\begin{align}
	\tilde{\Psi}(E,\eta)=\exp\left(- v^2\int d^3x\,{\rm Tr}\big(\Lambda^\dagger\Lambda \big)\right)\label{vortexansatz}\;.
\end{align} 
From the explicit expression  of $\Lambda$ \eqref{lambda}, \eqref{LambdaTL} it is seen that for  large values of $v^2$ this wave functional is strongly peaked at $E=0$ and $\eta=0$. We can therefore also carry out the integrals over $E$ and $\eta$ in \eqref{p-i} in the (lowest order) saddle point approximation, which then yields for the Wilson loop \eqref{p-i}
\begin{align}
	\langle W(C)\rangle=\sum_k \, \tilde{\Psi}\left(\frac{\mathscr{C}_k}{4\pi}\nabla \times \Sigma(S),-\frac{\mathscr{C}_k}{4\pi}\nabla\cdot \Sigma(S)\right)\;.
	\label{p-i1}
\end{align}
The remaining wave functional defines via Eq.  \eqref{infraredansatz} an effective theory in the presence of an external source given by the characteristic function $\Sigma(C)$ of the area enclosed by the Wilson loop $C$. Due to this source the dominant contribution to the functional integrals over the scalar fields comes here not from the vacuum solutions but from smooth classical domain wall solutions that interpolate between the different $ Z(N)$ vacua. Furthermore the domain walls are localized at the minimal surface $S_{\rm min}(C)$ bordered by the loop $C$. Note also that all co-weights $\mathscr{C}_k$ of the defining representation give the same contribution to the Wilson loop, so that the sum over them can be explicitly carried out. As detailed in Ref.  \cite{wavefunctional} one obtains then for asymptotically large loops $C$ an area law 
\begin{align}
     \langle W(C)\rangle \approx {\rm const}\times \exp\left[-\sigma A(S_{\rm min}(C))\right]\;,
     \label{WAlaw}
\end{align}
where $A(S_{\rm min}(C))$ is the area of the minimal surface $S_{\rm min}(C)$ bordered by the loop $C$ and $\sigma$ is the string tension of the defining representation. We also evaluated this average for a $k$-Antisymmetric representation and found that the string tension is compatible with a Casimir law when the parameters satisfy $\lambda a^2, \xi v^{N-2} >> \vartheta$.

\section{The 't Hooft loop}

According to Ref. \cite{thooft}, the algebraic properties (commutators) of Wilson and ’t Hooft loop operators lead to an area/perimeter complementarity when their expectation values are taken in the ground state of the theory. However, these properties do not by themselves enforce
that an area law for the average of one operator implies a perimeter law for the other average, when considering a given generic quantum state different from the ground state.
For this reason, it is essential to subject our wave functional to the 't Hooft loop test. If the wave functional peaked at center vortices we proposed has a good overlap with the vacuum in the IR regime, then it must display
this complementarity.

In the following we calculate the expectation value of the 't Hooft loop operator \eqref{thooftexplicit} in our vortex wave functional using both the explicit vortex representation. \eqref{wave-funct} and the effective field theory representation \eqref{infraredansatz}.

\subsection{Vortex Representation}
\label{sec:simple}

Before addressing the 't Hooft loop average by using the effective field representation of our wave functional, we shall consider its vortex representation, where the vortex guiding centers appear explicitly. Here, the argument we present disregards the interactions between the vortex lines, necessary to stabilize the condensate. They are incorporated in the next section, within the effective field representation. Under this condition, it is easy to see that our vortex wave functional \eqref{wave-funct} yields indeed a perimeter law for the 't Hooft loop. For this purpose we use the representation \eqref{wave-funct} for the wave functional and apply Eq.  \eqref{vort-gen} to obtain
\begin{align}
	\label{wave-funct1}
	V(C)&\Psi(A)=\Psi(A+a(C)) \nonumber \\
	&= \sum_{\{\gamma\}}\psi_{\{\gamma\}}\delta(A- a(\{\gamma\})+a(C)) \;,
\end{align}
Using the relation $a(-C)=-a(C)$ and defining the vortex configuration
\begin{equation}
	\{\gamma'\}=\{\gamma\}\cup (-C)\, ,
\end{equation} 
implying
\begin{equation}
	a(\{\gamma'\})=a(\{\gamma\})+a(-C)\, .
\end{equation}
Then, using the property of the statistical weight \eqref{stat-weight}:
\begin{equation}
	\psi_{\{\gamma'\}}=\psi_{\{\gamma\}}\psi_{-C} \,,
\end{equation}
we obtain
\begin{align}
	\label{wave-funct2}
	V(C)\Psi(A)&=\psi_{-C}^{-1} \sum_{\{\gamma'\}}\psi_{\{\gamma'\}}\delta(A- a(\{\gamma'\}))
	\nonumber \\
	&=\psi_{-C}^{-1}\Psi(A) \;,
\end{align}
With this result we find for the expectation value of the 't Hooft loop operator
\begin{equation}
	\label{stat-weight-l}
	\langle V(C)\rangle =\psi_{C}^{-1}\, ,
\end{equation}
where we have used that the statistical weight  \eqref{stat-weight} does not depend on the orientation of the vortex line, $\psi_{-C}=\psi_{C}$.

In the statistical weight $\psi_\gamma$ \eqref{stat-weight} the parameter $\mu$ controls the vortex density, and for percolating vortices, required for confinement, $\mu<0$ holds. Now, consider for simplicity a circular loop with radius $R$, which we put in the center of the $x-y-$ plane. In radial coordinates the length parameter is $s=R\varphi$ and the curvature of the circle is $\dot{\vu}(s)^2=\ddot{\vx}(s)^2=1/R^2=const$. Then from Eq.  \eqref{stat-weight-l} we find for the 't Hooft loop 
 \begin{equation}
 	\label{hooft-loop-final}
 	\langle V(C)\rangle =\exp{\bigg[-L\Big(|\mu|-\frac{1}{\kappa R^2}\Big)\bigg]}\, , \qquad L=2\pi R \, .
 \end{equation}
 For $R\to \infty$ the second term in  the exponent vanishes and we are left with a pure perimeter law with the perimeter tension $|\mu|$ exclusively determined by the vortex density.

\subsection{Effective Field Theory Representation}

The physical applicability of the wave functional we proposed in Ref.~\cite{wavefunctional}, describing an ensemble of center vortices, is restricted to the description of asymptotic IR properties of the Yang--Mills theory. In addition, as in any framework that goes from microscopic degrees of freedom---in this case, the thin center-vortex guiding centers---to macroscopic coarse-grained field variables, the effective representation can be trusted only at large distances.
Indeed, in the ensemble superposition, there is a series of approximations leading to the representation in Eqs.~\eqref{infraredansatz}--\eqref{fieldtheory}. In particular, its quadratic character in the field derivatives arises from the combined effect of stiffness and the assumption of sufficiently long center-vortex lines. Under these conditions, their end-to-end probabilities are well approximated by a second-order Fokker--Planck equation. In other words, the information that can be reliably obtained from this representation is entirely of infrared origin, i.e. below an appropriate momentum scale $\mathcal{Q}$. As we have seen, the calculation of the asymptotic Wilson-loop average is due to a smooth soliton and therefore it is a reliable low-momentum effect.

In the gauge-field representation, $V(C)$ \eqref{thooft-op} translates the argument of a state by the field of a center-vortex, see eq. \eqref{vort-gen}. In the electric field representation, where $\Pi(x)=-E(x)$, the operator of the 't Hooft loop \eqref{thooftexplicit} is diagonal and just multiplies the wave functional by a phase:
\begin{align}
	V(C)\tilde{\Psi}(E,\eta)=\exp\left[-i\int d^3x\, E \cdot a(C)\right]\tilde{\Psi}(E, \eta)
	\;.\label{thooft-expr}
\end{align}
With this relation we find for the 't Hooft loop 
\begin{align}
	&\langle V(C)\rangle = \int [DE][D \eta]\,  \tilde{\Psi}^*(E, \eta)e^{-i\int d^3x\, E \cdot a(C)} \tilde{\Psi}(E, \eta).
	\label{thooft-gen1} 
\end{align} 
Using the explicit expressions for the center vortex field $ a(C)$ \eqref{a2} and for $\Lambda$ \eqref{LambdaTL} we have
\begin{align}
	\int d^3x\, E \cdot  a(C)=-\frac{1}{4\pi}\int d^3x\, \Lambda^T(E)\cdot j(C)\;.\label{only-t}
\end{align}
Since the wave functional \eqref{infraredansatz} depends on the dual variables $E, \eta$ only through $\Lambda$, $\tilde{\Psi}(E, \eta)\equiv\tilde{\Psi}(\Lambda(E, \eta))$ it is convenient  to change the integration variable from $E(x)$ and $\eta(x)$ to $\Lambda^T(x)$ and $\Lambda^L(x)$, which can be easily done due to the linear relations \eqref{LambdaTL} (the Jacobian is an irrelevant constant). This yields for the 't Hooft loop
\begin{align}
	\langle V(C)\rangle  = \int [D\Lambda]\,	\tilde{\Psi}^*(\Lambda)e^{\frac{i}{4\pi}\int d^3x\, \Lambda\cdot j(C)} \tilde{\Psi}(\Lambda).
	\label{thooft-loop-e}
\end{align}
The wave functional $\tilde{\Psi}(\Lambda)=\tilde{\Psi}(E,\eta)$ is given by the effective field theory \eqref{infraredansatz}. Moreover, due to Eq. \eqref{only-t} only the transverse part of $\Lambda$ is relevant for the calculation of the 't Hooft loop. In what follows, we therefore retain only $\Lambda^T$, but omit the superscript for notational simplicity. We evaluate this field theory consistent with the caculation of the Wilson loop given in sect. \ref{WL-sect}: First we carry out the functional integral over the scalar field $\Phi$ in $\tilde{\Psi}^*(\Lambda)$. We will do this at one loop level, by considering an expansion of the scalar fields around their vacuum and retaining only the quadratic part of the action. After integration over the fluctuations, this leads to (see Appendix \ref{app-1}) 
\begin{gather}
   \tilde{\Psi}^*(\Lambda)=\exp\left[-\int d^3x\left(\frac{v^2}{2}\Lambda^2+\frac{F^2(\Lambda)}{2m}\right)\right]\;.
\end{gather}
Then the 't Hooft loop is given by
\begin{gather}
    \langle V(C)\rangle \approx \int [D\Lambda]\exp\left[-\int d^3x\left(\frac{v^2}{2}\Lambda^2+\frac{F^2(\Lambda)}{2m}\right)\right]\times\nonumber\\\exp\left[\int d^3x \frac{i}{4\pi}\Lambda\cdot j(C)\right] \tilde{\Psi}(\Lambda)\;.
\end{gather}
As the first factor is a Gaussian, we approximate the integral by the value of the integrand on its peak, given by
\begin{equation}
    \Lambda_0(C)=\frac{i}{4\pi}O^{-1}j(C)\;,O=\frac{-\nabla^2+M^2}{m}\;, M^2=v^2m\;. \label{ext-field}
\end{equation}
Then, the 't Hooft loop is
\begin{gather}
    \langle V(C)\rangle =e^{\frac{i}{8\pi}\int d^3x \Lambda_0(C)\cdot j(C)}\tilde{\Psi}(\Lambda_0(C))\;.\label{thooft-gen3} 
\end{gather}
The terms in the exponent are highly localized on the curve $C$, so that they are expected to yield a contribution proportional to the perimeter of the curve. The factor $\tilde{\Psi}(\Lambda_0(C)) $ is still a field theory, which we will treat again in the saddle-point approximation in the next section.

\section{The 't Hooft Loop Soliton}

Here, we shall study the contribution to the ’t Hooft loop $\langle V(C)\rangle$ originating from $\tilde{\Psi}(\Lambda_0(C))$ in Eq. \eqref{thooft-gen3}, by means of a saddle-point approximation of its scalar-field representation. Due to the localized nature of the ''external source" $\Lambda_0(C)$ we expect that the saddle-point of the scalar field $\Phi$ will be given by a soliton configuration which is strongly localized at the source $\Lambda_0(C)$ and approaches the vacuum value $\Phi=vI$ far away from the loop $C$. Indeed, in that region, the external source $\Lambda_0(C)$ is exponentially suppressed at scales larger than $1/M$. In addition, the action functional $W(\Phi,0)$ has discrete vacua and all the associated scalar field-modes are massive. This is a consequence of having a condensate of center-vortices with nonoriented components. Therefore, the minimization of $W(\Phi,\Lambda_0(C))$ results in soliton fields that approach one of these vacua far away from the loop $C$, which is a connected region.  Thus, on general grounds, it is clear that the action of the soliton field will be concentrated around $C$, giving rise to a perimeter law.    

It is interesting to compare this situation with that obtained when computing the Wilson loop. As shown in Ref. \cite{wavefunctional}, and briefly reviewed in Sec. \ref{WL-sect}, in that case the action minimization involves boundary conditions on a disconnected set, with different vacua in the regions with large positive and negative coordinates transverse to the planar loop. A domain wall sitting on the Wilson loop was thus obtained, generating a minimal area law. It is satisfying to see that the area/perimeter complementarity emerges as a clear consequence of the very same mechanism. At the level of the IR scalar field representation of the wave functional, the mixed center-vortex condensate produces gapped field modes and discrete vacua. This, in turn, generates a large-distance behavior of observables that depend on the connectivity properties of the vacuum regions induced by the corresponding sources.

An asymptotically large circular loop looks locally like a straight line. We can therefore expect that the perimeter coefficient of the 't Hooft loop $ \langle V(C)\rangle $ can be estimated by considering the limiting case where $C$ becomes an infinite straight line. Locating this line on the $z$-axis the external field $\Lambda_0(C)$ \eqref{ext-field} in the argument of the wave functional in Eq. \eqref{thooft-gen3} is given by
\begin{equation}
	\Lambda_0(x)=iO^{-1} j(x), \qquad j(x)=\mathscr{C} \delta^2(x_\perp) \ve_z
\end{equation}
where $x_\perp$ denotes a vector in the plane perpendicular to the $z$-axis. 
Due to the axial symmetry of this field it is convenient to use cylindrical coordinates $x=(\rho, \varphi,z)$. In these coordinates the external 
\begin{gather}
    \Lambda_0=i\mathscr{C}\frac{m}{8\pi^2}K_0(M\rho)\ve_z\;,
\end{gather}
where $K_0(x)$ is the modified Bessel function of the second kind. We note that this field is imaginary, so that we need to extend our wave functional, initially defined for real $\Lambda$, to the imaginary axis. We do this continuation such that the effective field theory \eqref{action-sf} representing our wavefunctional remains well-defined. For the exponent in Eq. \eqref{thooft-gen3}, using the relation
\begin{gather}
(\mathscr{C},\mathscr{C})=8\pi^2\frac{N-1}{N}\;, 
\end{gather}
we obtain
\begin{gather}
    i\int d^3x \Lambda_0\cdot j(C) = -\frac{m(N-1)}{2\pi N}\int d\rho K_0(M\rho)\delta(\rho)\int dz\;,
\end{gather}
which yields a divergent contribution to the perimeter law. The divergence proportional to the perimeter is expected \cite{polyakov}, and can be dealt with by means of a renormalization of the 't Hooft operator. Then, we may write
 \begin{align}
     \langle V_C\rangle =e^{-\gamma_d L(C)}V_{\rm ren}(C)\;,
 \end{align}
 where $L(C)$ is the perimeter of $C$, and $\gamma_d$ is a divergent constant. In fact such a singular behaviour of the perimeter term is also observed in the lattice calculation of the 't Hooft loop \cite{polyakov}. In the following we will show that the second factor in Eq.  \eqref{thooft-gen3}, which contributes to the renormalized 't Hooft operator, will indeed give rise to a perimeter law with a finite coefficient.

\subsection{The Soliton Solution}

\begin{figure}[htbp]
    \centering
        \includegraphics[scale=0.6]{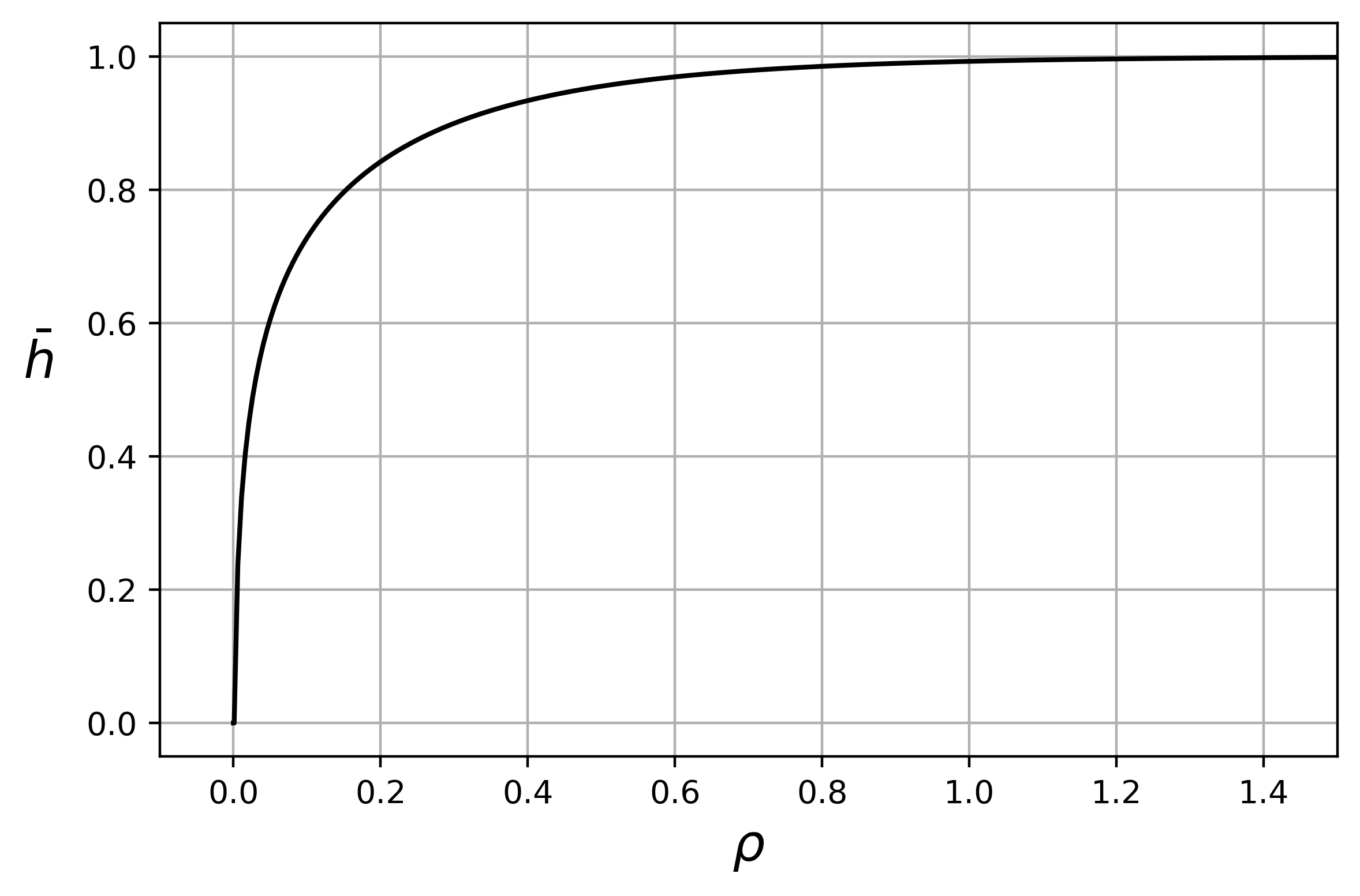}
    \caption{\justifying Dimensionless radial profile $\bar{h}=h/v$ for $ \lambda=40, \xi=1, \vartheta=0.01, a=0.3$. In this case, the contribution to $\gamma$ is $0.24$. }
    \label{fig:solit-infty}
\end{figure}

As discussed above, a nontrivial soliton solution of the field theory defining our wave functional $\tilde{\Psi} (\Lambda_0)$ is expected due to the localized form of the external field $\Lambda_0$ induced by the 't Hooft loop.
Furthermore uniqueness of the solution requires
\begin{equation}
	\label{bc2}
	\Phi (\rho, \varphi +2\pi,z)=	\Phi(\rho, \varphi,z).
\end{equation}
As $\Lambda_0\cdot \nabla \Phi=0$, the equation of motion simplifies to
\begin{align}
	\nabla^2\Phi&+\Lambda_0^\dagger\Lambda_0\Phi = \lambda \Phi(\Phi^\dagger\Phi-a^2) \nonumber \\
	&-\xi\det \Phi^*(\Phi^*)^{-1}-2\vartheta T_A \Phi T_A\;.
	\label{eomm}
\end{align}
This equation confirms our initial expectations. Far from the loop, the source vanishes and the fields remain at their vacuum value $v$. As the loop is approached, however, the source induces an effective positive mass squared term, breaking the dual superconducting phase and giving rise to a finite energy density localized on the loop. It is sufficient to solve this equation for $SU(2)$ since the higher gauge groups $SU(N>2)$ contain the $SU(2)$ as subgroup and, as in the case of the instantons, we can expect that the embedding of the $SU(2)$ soliton into $SU(N)$ gives the lowest $SU(N)$ soliton solution.  In this case, the external source $\Lambda_0$ induced by the 't Hooft loop via the saddle-point condition reads
\begin{gather}
    \Lambda_0 = i\frac{m}{4\pi} K_0(M\rho) \frac{\tau_3}{2}\ve_z\;.
\end{gather}
We shall assume that the soliton solutions have the form:
\begin{equation}
	\label{sol-f1}
	\Phi=h u
\end{equation}
where $h$ is a real scalar field and 
\begin{equation}
	\label{chiral-f}
	u=\exp{(i\chi \tau_3)  }=\cos\gamma+i\tau_3\sin\gamma
\end{equation}
(with $\chi$ being real) is a chiral field living in the Cartan subgroup $U(1)$ . With this representation and $T_a=\tau_a/2$ the equation of motion \eqref{eomm} contains two independent equations given by the terms proportional to the unit matrix and to $\tau_3$, respectively, which read:
\begin{align}
    &\cos \chi(\nabla^2h-h (\nabla \chi)^2)-\sin \chi (h \nabla^2 \chi + 2 \nabla \chi\cdot \nabla h)+\nonumber\\& \frac{m^2}{64\pi^2} K_0^2(M\rho) h=\cos\chi (\lambda h(h^2-a^2)-\left(\frac{3}{2}\vartheta+\xi\right)h)\;,\nonumber\\&
    \cos \chi(h\nabla^2\chi+2\nabla\chi\cdot\nabla h)+\sin\chi(\nabla^2h-h(\nabla\chi)^2)=\nonumber\\&\sin\chi(\lambda h(h^2-a^2)-\left(\xi-\frac{\vartheta}{2}\right)h)\;.
\end{align}
As $\chi$ is not coupled to the external source, we expect that the lowest energy solution will correspond to $\gamma=0$. Then, the second equation is satisfied, and the first one simplifies to
\begin{align}
    \nabla^2h+\frac{m^2}{64\pi^2} K_0^2(M\rho) h=\lambda h (h^2-v^2)\;, \label{our-eq}
\end{align}
where the vacuum value of the scalar field $v$ is 
\begin{gather} v^2=a^2+\frac{2\xi+3\vartheta}{2\lambda}\;.
\end{gather}
The energy per unit length of the solution is
\begin{align}
    \gamma &= 2\int d^2x \left((\nabla h)^2+\frac{m^2}{64\pi^2} K_0^2(M\rho) h^2\right)\nonumber\\&+\int d^2x \left(\lambda (h^2-a^2)^2-2\xi h^2 -3\vartheta h^2\right)\;.
\end{align}
 Moreover, in order to obtain a finite energy, the scalar profile has to approach the vacuum as $\rho \to \infty$, i.e. $h(\rho\to\infty)=v$. To obtain a well-defined solution, we must require \(v^2 > 0\), which corresponds to the spontaneously broken (SSB) phase of the scalar field theory with discrete $Z(N)$ vacua. This condition generalizes the earlier requirement \(\mu < 0\) found in Eq. ~\eqref{hooft-loop-final}, where interactions between vortex lines were neglected. Once these interactions are included, what we actually need is the formation of a condensate—precisely the same condition required for the area law of the Wilson loop.  We have solved Eq. \eqref{our-eq} subject to a Dirichlet boundary condition at $\rho=0$. In fig. \ref{fig:solit-infty} we plot the dimensionless profile $\bar{h}=h/v$ for a parameter choice compatible with our condition $\lambda a^2, \xi >> \vartheta$ for Casimir scaling. The localization scale of the source for this choice of parameters is $1/M \approx 1/8$ (see Eq. \eqref{ext-field}). We see that the scalar field performs a smooth transition from the trivial configuration $\bar{h}=0$ to one of the $Z(N)$ vacua. This confirms our expectations of a finite contribution to the perimeter law as due to a smooth solution whose energy density is localized near the loop.

\section{Summary and Conclusions}

In Ref. \cite{wavefunctional}, we proposed a wave functional peaked at a center-vortex condensate, which can be represented in terms of a scalar field theory living on a 3D equal-time slice. Because of the presence of not only oriented center vortices but also nonoriented ones, the scalar modes turned out to be gapped and the effective field representation displays discrete $Z(N)$ vacua. This, together with the fact that the calculation of the Wilson loop involved boundary conditions on disconnected regions, led to a domain wall soliton solution localized on the minimal area bordered by the Wilson loop, thereby producing an area law.

In the present work we used the center vortex wavefunctional proposed in Ref. \cite{wavefunctional} to calculate the spatial 't Hooft loop and found a perimeter law. For a sufficiently large circular loop $C$, so that the effective IR field representation of our wavefunctional applies, the dominant field configurations are solitons that approach one of the discrete vacua outside a toroidal neighborhood of $C$, which is a connected region. Therefore, on general grounds, the relevant soliton configurations have action localized around $C$, giving rise to a perimeter law. The area/perimeter complementarity thus follows from the same underlying mechanism: the mixed center-vortex condensate generates gapped modes and discrete vacua, and the large-distance behavior of observables is governed by the connectivity of the vacuum regions selected by the external sources induced by the Wilson or 't Hooft loop, respectively. As a consequence, the proposed IR wave functional satisfies both confinement criteria.

\section{Acknowledgments}
The authors are grateful to D. Campagnari for useful discussions. This work was financed, in part, by the S\~ao Paulo Research Foundation (FAPESP), Process No. 2023/18483-0 (D.R.J.), No. 2024/20896-3 (D.R.J). Financial support from the Brazilian agency CNPq under Contract No. 309971/2021-7 (L.E.O.) is also
gratefully acknowledged.
\appendix
\section{Effective field representation for the wave functional}
\label{app-0}
The equivalence between the representations \eqref{we}, \eqref{infraredansatz} for our wavefunctional can be seen by expanding \eqref{infraredansatz} in powers of $\xi_0$ and $\vartheta_0$. To lowest nontrivial order, the $\xi_0$ expansion yields
\begin{align}
    \xi_0^2 \prod_{k=1}^NG_k(x,y) \det O_{\mathscr{C}_{k}}\makebox[.5in]{,}O_{\mathscr{C}_{k}}=D^2(\Lambda_k)+\mu-i\sigma \label{n-match}\;,
\end{align}
where $G_k(x,y)=\langle\phi_k(x)\bar{\phi}_k(y)\rangle$ is the two-point function of the field $\phi_k$. To interpret this contribution, it is convenient to define $Q_{\mathscr{C}}(x,u,x_0,u_0,L)$, the partial sum over vortex lines $l$ with length L carrying a co-weight $\mathscr{C}$, which begin at $x_0$ with an orientation $u_0$ and end at $x$ with an orientation $u$, as follows
\begin{gather}
    Q_{\mathscr{C}}(x,u,x_0,u_0,L)=\int [dx]^L_{v_0,v}\psi_l e^{i \int_l dx\cdot \Lambda_{\mathscr{C}}}\;,
\end{gather}
where we have denoted the loop coordinates and orientation by $v\equiv (x,u)$, and the measure $[dx]^{L}_{v_0,v}$ integrates over the vortex lines satisfying the above-mentioned conditions. As discussed in Refs. \cite{dgo}, \cite{ox-reinhardt}, \cite{ALB}, this object satisfies the diffusion equation 
\begin{align}
    -\partial_L Q_{\mathscr{C}}= \left(-\frac{\kappa}{2}\Delta_u+\mu-i\sigma +u\cdot (\nabla-i\Lambda_{\mathscr{C}})\right)Q_{\mathscr{C}}\;,
\end{align}
where $\Delta_u$ is the Laplacian on the unit sphere. In the limit of small stiffness (large $\kappa$), the partial amplitude becomes independent of the orientations, and the equation becomes approximately
\begin{align}
    -\partial_L Q_{\mathscr{C}}(x,x_0,L) \approx O_{\mathscr{C}}Q_{\mathscr{C}}\;.
\end{align}
In this limit, the solution is thus
\begin{align}
    Q_{\mathscr{C}}(x,x_0,L)\approx \langle x | e^{- L O_{\mathscr{C}}}|x_0\rangle \;.\label{build-block-1}
\end{align}
The contribution of all loops with the given co-weight can be written as
\begin{align}
   &\tilde{\Psi}_{\mathscr{C}}(E,\eta) =\sum_{n=0}^\infty \frac{1}{n!}\prod_{i=1}^n\int \frac{dL}{L}\int dv_i\int[dx^{(i)}]^{L_i}_{v_i,v_i}\psi_{\gamma_i}\times\nonumber\\&e^{i \int_{\gamma_i} dx\cdot \Lambda_{\mathscr{C}}}\;.
\end{align}
The sum over loops can be exponentiated in terms of $\mathcal{Q}_\mathscr{C}$ as follows
\begin{equation}
   \tilde{\Psi}_{\mathscr{C}}(E,\eta)=e^{-\int \frac{dL}{L}{\rm tr}Q_{\mathscr{C}}(x,x,L)}\;.  
\end{equation}
Then, using Eq. \eqref{build-block-1}, we obtain
\begin{align}
\tilde{\Psi}_{\mathscr{C}}(E,\eta) =\det O_{\mathscr{C}}\;.
\end{align}
 That is, the factor $\det O_{\mathscr{C}_k}$ in Eq. \eqref{n-match} generates the contribution of the closed loops carrying the co-weight $\mathscr{C}_k$. We find for the Green's function in Eq. \eqref{n-match}, the representation
\begin{align}
    &G_{\mathscr{C}}(x,y)=\int dL Q_{\mathscr{C}}(x,x_0,L)\;,
\end{align}
by integrating Eq. \eqref{build-block-1} over $L$. Therefore, 
\begin{align}
    \xi_0^2 \prod_{k=1}^NG_k(x,y) = \psi_{\{\gamma\}}\prod_{k=1}^Ne^{i \int_{\gamma_k}dx \cdot \Lambda_{\mathscr{C}_{k}}}\;,
\end{align}
where $\{\gamma\}$ is the "N-matching" vortex network, formed by $N$ vortex lines with common endpoints carrying the $N$ possible co-weights. Similarly, the expansion of \eqref{infraredansatz} to lowest order in $\vartheta_0$ yields
\begin{align}
    \sum_{i\neq j}\vartheta_0^2 G_i(x,y)G_j(y,x) \det O_{\mathscr{C}_i} \det O_{\mathscr{C}_j}\;, 
\end{align}
which gives the contribution of a network containing two vortex lines interpolated by monopoles, in addition to the contribution due to the closed vortex loops carrying co-weights $\mathscr{C}_{i}, \mathscr{C}_{j}$. Then, when all powers of $\xi_0, \vartheta_0$ are considered, an ensemble of vortex loops, together with all possible $N$-matching and non-oriented configurations, is generated.
\section{Integrating over fluctuations of the scalar fields}\label{app-1}
Our wavefunctional is given by the scalar field theory \eqref{action-sf}. In the following we work out this field theory in the semiclassical approximation, i.e. expanding the field around its vacuum value and performing the Gaussian integral over the fluctuations. For simplicity we assume the $SU(2)$ gauge group. Then, the potential becomes
\begin{align}
    &V(\phi)=\frac{\lambda}{2}(|\phi_1|^4+|\phi_2|^4)-\left(\frac{\vartheta}{2}+\lambda a^2\right)(|\phi_1|^2+|\phi_2|^2)\nonumber\\&-\xi(\phi_1\phi_2+\bar{\phi}_1\bar{\phi}_2)-\vartheta(\phi_1\bar{\phi}_2+\bar{\phi}_1\phi_2)\;.
\end{align}
The vacua of this potential are given by $\phi_1=\phi_2=v$, where
\begin{gather}
v^2=a^2+\frac{2\xi+3\vartheta}{2\lambda}\;.
\end{gather}
Consider fluctuations around this vacua, parametrized by complex functions $\varphi_1,\varphi_2$:
\begin{align}
    \phi_1=v+\varphi_1\makebox[.5in]{,}\phi_2=v+\varphi_2\;.
\end{align}
The expansion of the potential around the vacuum, up to quadratic order, reads
\begin{gather}
    V=\frac{\lambda v^2}{2}(\varphi_1^2+\varphi_2^2+\bar{\varphi}_1^2+\bar{\varphi}_2^2)+\nonumber\\\left(\lambda(2v^2-a^2)-\frac{\vartheta}{2}\right)(|\varphi_1|^2+|\varphi_2|^2)-\xi(\varphi_1\varphi_2+\bar{\varphi}_1\bar{\varphi}_2)\nonumber\\-\vartheta(\varphi_1\bar{\varphi}_2+\bar{\varphi}_1\varphi_2)\;.
\end{gather}
Defining the vector
\begin{gather}
    \varphi =\begin{pmatrix}
        \varphi_1\\
        \varphi_2 \\
        \bar{\varphi}_1\\
        \bar{\varphi}_2
    \end{pmatrix}\;,
\end{gather}
the potential can be written as
\begin{align}
    V=\varphi^\dagger\mathcal{M}^2 \varphi\;.
\end{align}
where the matrix $\mathcal{M}^2$ is
\begin{gather}
   \mathcal{M}^2=\frac{1}{2}\begin{pmatrix}
        C&B\\
        B&C
    \end{pmatrix}\;\;,
\end{gather}
where
\begin{gather}
C=\begin{pmatrix}
        \lambda(2v^2-a^2)-\frac{\vartheta}{2}&-\vartheta\\
        -\vartheta& \lambda(2v^2-a^2)-\frac{\vartheta}{2}\end{pmatrix}\;,\nonumber\\ B=\begin{pmatrix}
            \lambda v^2&-\xi \\-\xi&\lambda v^2
        \end{pmatrix}\;.
\end{gather}

Up to second order in the fluctuations the action becomes
\begin{gather}
    W=\int d^3x \left(\varphi^\dagger(N(\Lambda)+\mathcal{M}^2)\varphi+iJ^\dagger\varphi\right) \label{firstint}\;,\\
    N(\Lambda)=\frac{1}{2}\begin{pmatrix}
        -D^2(\Lambda)I_{2\times 2}& 0\\
        0 & -D^2(\Lambda)I_{2\times 2}
    \end{pmatrix}\;,\nonumber\\J=\begin{pmatrix}-\omega_1\cdot(\nabla\cdot\Lambda)+v(\Lambda\cdot\omega_1)^2\\-\omega_2\cdot(\nabla\cdot\Lambda)+v(\Lambda\cdot\omega_2)^2\\ \omega_1\cdot(\nabla\cdot\Lambda)+v(\Lambda\cdot\omega_1)^2\\ \omega_2\cdot(\nabla\cdot\Lambda)+v(\Lambda\cdot\omega_2)^2)\end{pmatrix}\;.
\end{gather}
Integration over $\varphi, \varphi^\dagger$ leads to 
\begin{gather}
    \int [D\varphi^\dagger][D\varphi]e^{-W(\varphi,\varphi^\dagger)}=\det(-D^2+C+B)^{-\frac{1}{2}}\times\nonumber\\ \det(-D^2+C-B)^{-\frac{1}{2}}e^{-\frac{v^2}{2}\int d^3x \Lambda^2-\int d^3x J^\dagger (-N(\Lambda)+\mathcal{M}^2)^{-1}J}\label{detsu2}.
\end{gather}
Here, we have used that for a $2n\times 2n$ matrix $M$ of the type
\begin{gather}
    M=\begin{pmatrix}
        A &B\\B&A
    \end{pmatrix}\;,
\end{gather}
where $A,B$ are $n\times n$ matrices, the determinant is given by
\begin{gather}
    \det M =\det(A-B)\det(A+B)\;.
\end{gather}
This formula can be obtained by using the similarity transformation
\begin{equation}
    P=\begin{pmatrix}
        I &I\\I&-I
    \end{pmatrix}\makebox[.5in]{,}P^{-1}=\frac{1}{2}\begin{pmatrix}
        I &I\\I&-I
    \end{pmatrix}
\end{equation}
to diagonalize the matrix $M$. Following Ref. \cite{h-book}, we can evaluate the leading terms of the effective action for $\Lambda$ in a gradient expansion. This results is
\begin{gather}
   \exp\left[-\frac{F^2(\Lambda)}{2m}+\dots\right]\nonumber\;, \\m=192\pi \frac{(\vartheta+\xi)(4\vartheta+2 a^2\lambda+3\xi)}{5\vartheta+2a^2\lambda+4\xi}\;,
\end{gather}
where the dots represent terms containing higher-order derivatives of $\Lambda$.

\end{document}